\documentclass[12pt]{article}
\usepackage{graphicx}

\newcommand\pubnumber{WSU--HEP--XXYY}
\newcommand\pubdate{\today}

\def\wayne{Institute of High Energy Physics, Main Building B406 \\
 Chinese Academyof Science, Beijing, China}

\def\Title#1{\begin{center} {\Large #1 } \end{center}}
\def\Author#1{\begin{center}{ \sc #1} \end{center}}
\def\Address#1{\begin{center}{ \it #1} \end{center}}

\newcommand\pubblock{\rightline{\begin{tabular}{l} \pubnumber\\
         \pubdate  \end{tabular}}}
\newenvironment{Abstract}{\begin{quotation}  }{\end{quotation}}
\newenvironment{Presented}{\begin{quotation} \begin{center} 
             PRESENTED AT\end{center}\bigskip 
      \begin{center}\begin{large}}{\end{large}\end{center} \end{quotation}}
\def\Acknowledgements{\bigskip  \bigskip \begin{center} \begin{large}
             \bf ACKNOWLEDGEMENTS \end{large}\end{center}}

\def\beq{\begin{equation}}
\def\eeq#1{\label{#1}\end{equation}}
\def\eeqn{\end{equation}}

\def\beqa{\begin{eqnarray}}
\def\eeqa#1{\label{#1}\end{eqnarray}}
\def\eeqan{\end{eqnarray}}

\let\bar=\overbar

\def\Dslash{\not{\hbox{\kern-4pt $D$}}}
\def\dslash{\not{\hbox{\kern-2pt $\del$}}}

\def\msb{{\bar{\ssstyle M \kern -1pt S}}}

\begin{document}
\begin{titlepage}
\pubblock

\vfill
\Title{Semileptonic D-decays at BESIII}
\vfill
\Author{Fen-fen An}
\Address{\wayne}
\vfill
\begin{Abstract}
We present here three analyses of semileptonic $D$-meson decays based on the
2.92 fb$^{-1}$ of data collected by the BESIII experiment in 2010 
and 2011 at the $\psi$(3770) peak. For the decay
$D^{+}\to K^{-}\pi^{+}e^{+}\nu_{e}$, its branching fraction 
is measured over the whole $m_{K\pi}$ region and in the $\overline{K}^{*}(892)^{0}$
window, respectively. A partial wave analysis (PWA) is performed, 
indicating an \emph{S}-wave contribution of about 6\%.
The \emph{S}-wave phase and the form factors are measured 
 by the PWA
 and in a model-independent way. For the decay $D^{+}\to \omega e^{+}\nu_{e}$,
 an improved measurement of the branching fraction is performed
and the form factors are determined for the first time.
 $D^{+}\to \phi e^{+}\nu_{e}$ is searched and an improved upper limit 
 at 90\% confidence level is set. 
For the decay $D^{+}\to K_{L} e^{+}\nu_{e}$,
its branching fraction is measured for the first time
and the $CP$ asymmetry is presented. The product of the hadronic form 
factor and the CKM matrix element, $f_{+}^{K}(0)|V_{cs}|$,
is also determined in this decay.

\end{Abstract}
\vfill
\begin{Presented}
The 7th International Workshop on Charm Physics (CHARM 2015)\\
Detroit, MI, 18-22 May, 2015
\end{Presented}
\vfill
\end{titlepage}
\def\thefootnote{\fnsymbol{footnote}}
\setcounter{footnote}{0}
%

\section{Introduction}
\label{sec:introduction}

The analyses presented in this paper are based on the 2.92 fb$^{-1}$ of 
data sample~\cite{BESIII:2013iaa}
collected in $e^{+}e^{-}$ annihilations at the $\psi$(3770) peak,
which is accumulated with the BESIII detector~\cite{besiii} 
operated at the double-ring BEPCII collider. At the
$\psi$(3770) peak, $D$ mesons are produced in pairs, thus 
allowing us to use the technique of tagging $D$-meson decays~\cite{dtag method}.
If a decay of one $D$ meson (``tagged decay'')
has been fully reconstructed in an event, 
the existence of another $\bar{D}$ decay (``signal decay'')
in the same event is guaranteed. In these analyses 
$D^-$ is firstly reconstructed in one of the 
six decay channels:
$D^- \to K^+ \pi^- \pi^-$, 
$D^- \rightarrow K^+ \pi^- \pi^- \pi^0$, 
$D^- \rightarrow K_S^0 \pi^-$, 
$D^- \rightarrow K_S^0 \pi^- \pi^0$,
$D^- \rightarrow K_S^0 \pi^- \pi^- \pi^+$,
 and $D^- \rightarrow K^+ K^- \pi^-$,
 then semileptonic signals are searched  at the recoiling side.
 Charge conjugated decays are implied similarly.

\section{Study of $D^{+}\to K^{-}\pi^{+}e^{+}\nu_{e}$}
\label{sec:kpienu}

The invariant matrix element of the $D_{e4}$ decay of $D^{+}\to K^{-}\pi^{+}e^{+}\nu_{e}$ is 
the product of
a leptonic and a hadronic current. Thus it's convenient for us to study 
the $K\pi$ system in the absence of interactions with
other hadrons and  the hadronic transition form factors. 
In this analysis we measure the different $K\pi$ resonant 
and non-resonant amplitudes that 
contribute to this decay, which can provide helpful information for the \emph{B}-meson
semileptonic decays.
Also we measure the $q^{2}$ dependent transition form factors,
 where $q^{2}$ is the invariant mass square of
the lepton pair of the $e\nu_{e}$ system. This can be compared with  
hadronic model expectations and lattice QCD computations~\cite{Bernard:1991bz}.

The signal yield is obtained by simply counting 
since the background level is lower than 1\%.
Using the tagging technique introduced in Sec.~\ref{sec:introduction},
the branching fractions are measured to be $(3.71\pm0.03\pm0.08)$\%
over the full $m_{K\pi}$ range and $(3.33\pm0.03\pm0.07)$\%
in the $\overline{K}^{*}(892)^{0}$ region, respectively.

The 4-body $D_{e4}$ decay  can be uniquely described 
by five kinematic variables~\cite{pwa5}: 
$K\pi$ mass square ($m^{2}$), $e\nu_{e}$ mass square ($q^{2}$), the angle 
between the $\pi$ and the $D$ direction in the $K\pi$ rest frame 
($\theta_{K}$), the angle between the $\nu_{e}$ and the $D$ 
direction in the $e\nu_{e}$ rest frame 
($\theta_{e}$), and the angle between the two decay planes ($\chi$).
Its PDF can be expressed based on the five variables and 
then the PWA is performed using an unbinned maximum 
likelihood method. 

\begin{figure}[h]  
  \begin{center}
  \includegraphics[width=0.28\linewidth]{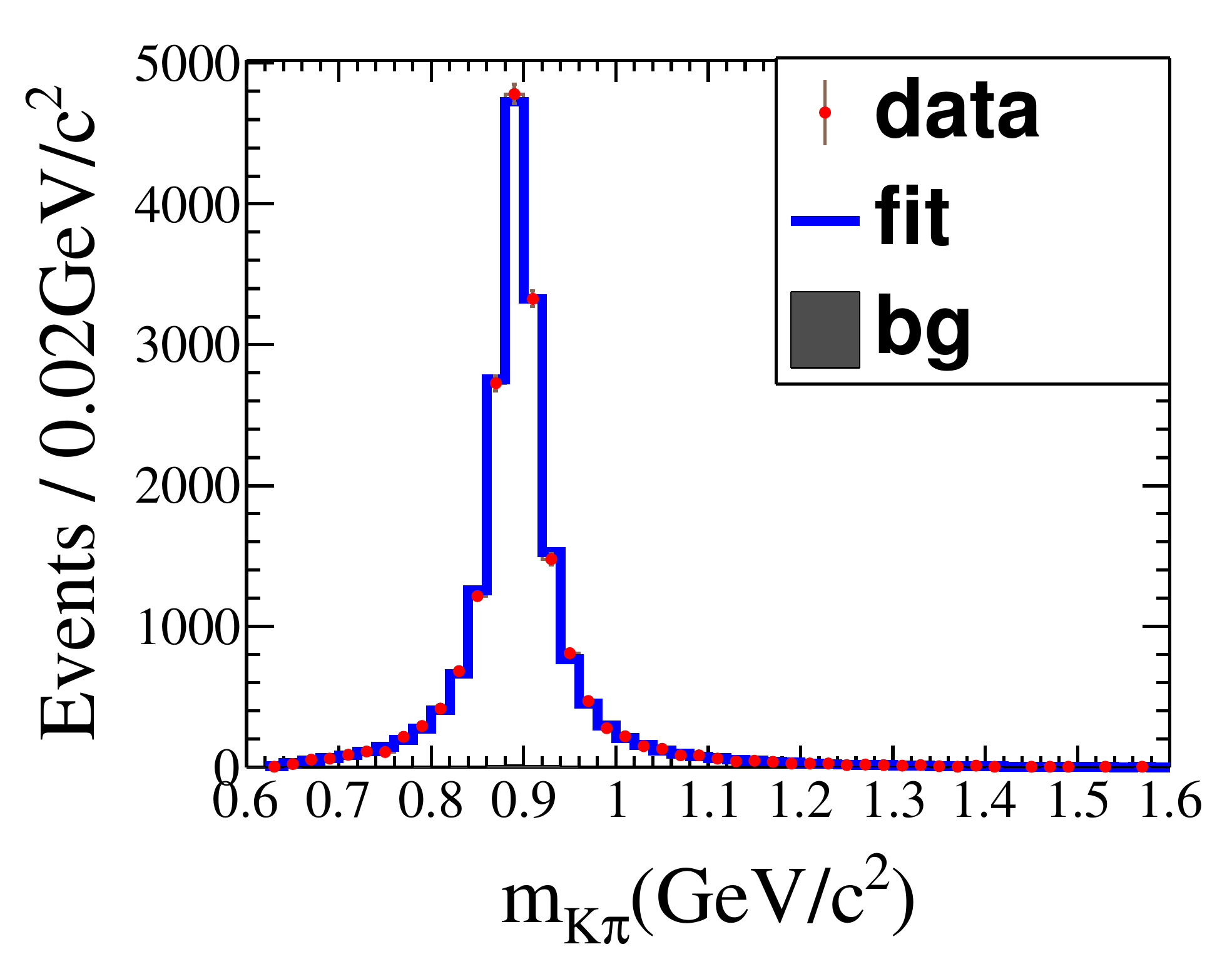}
  \includegraphics[width=0.28\linewidth]{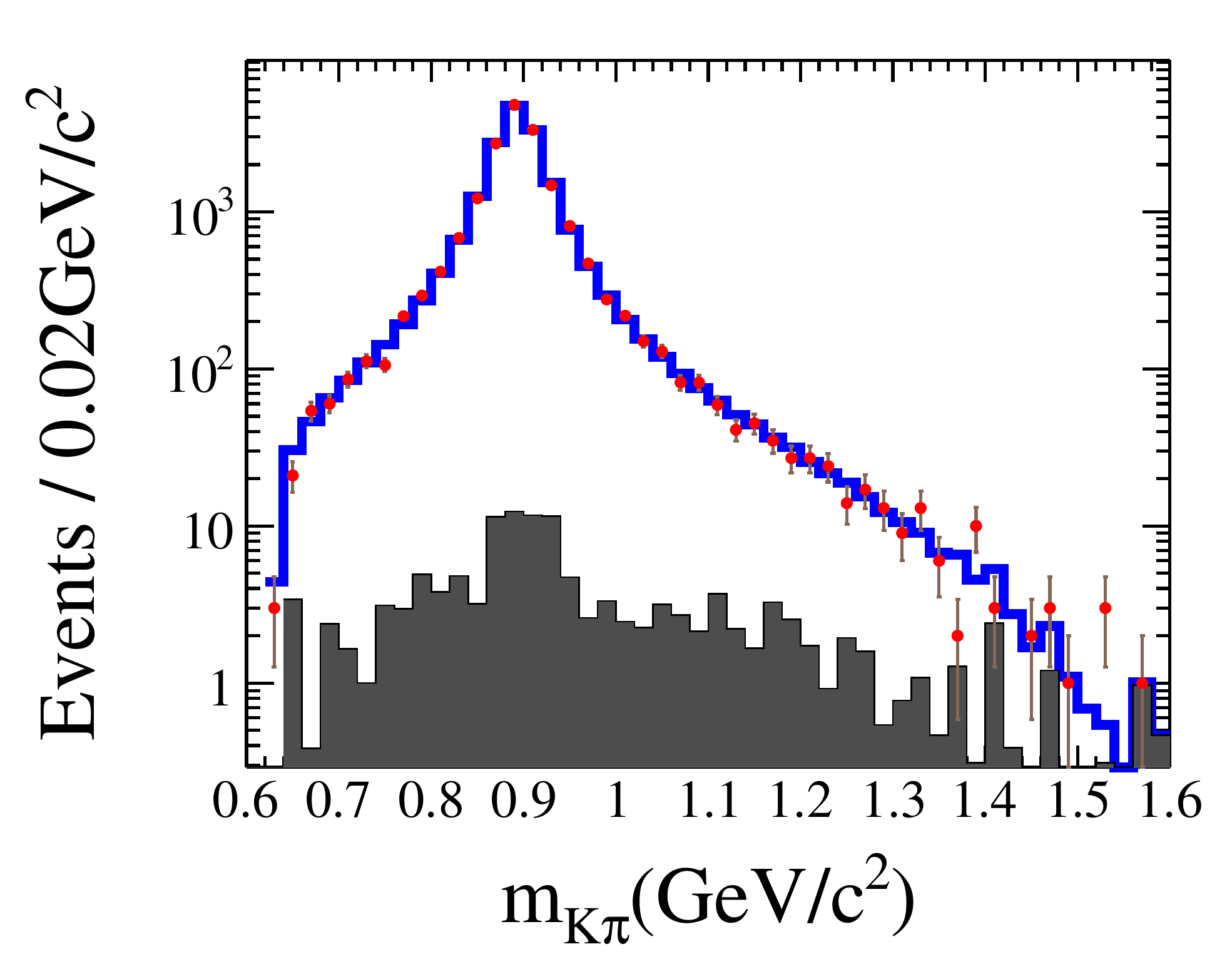} 
  \includegraphics[width=0.28\linewidth]{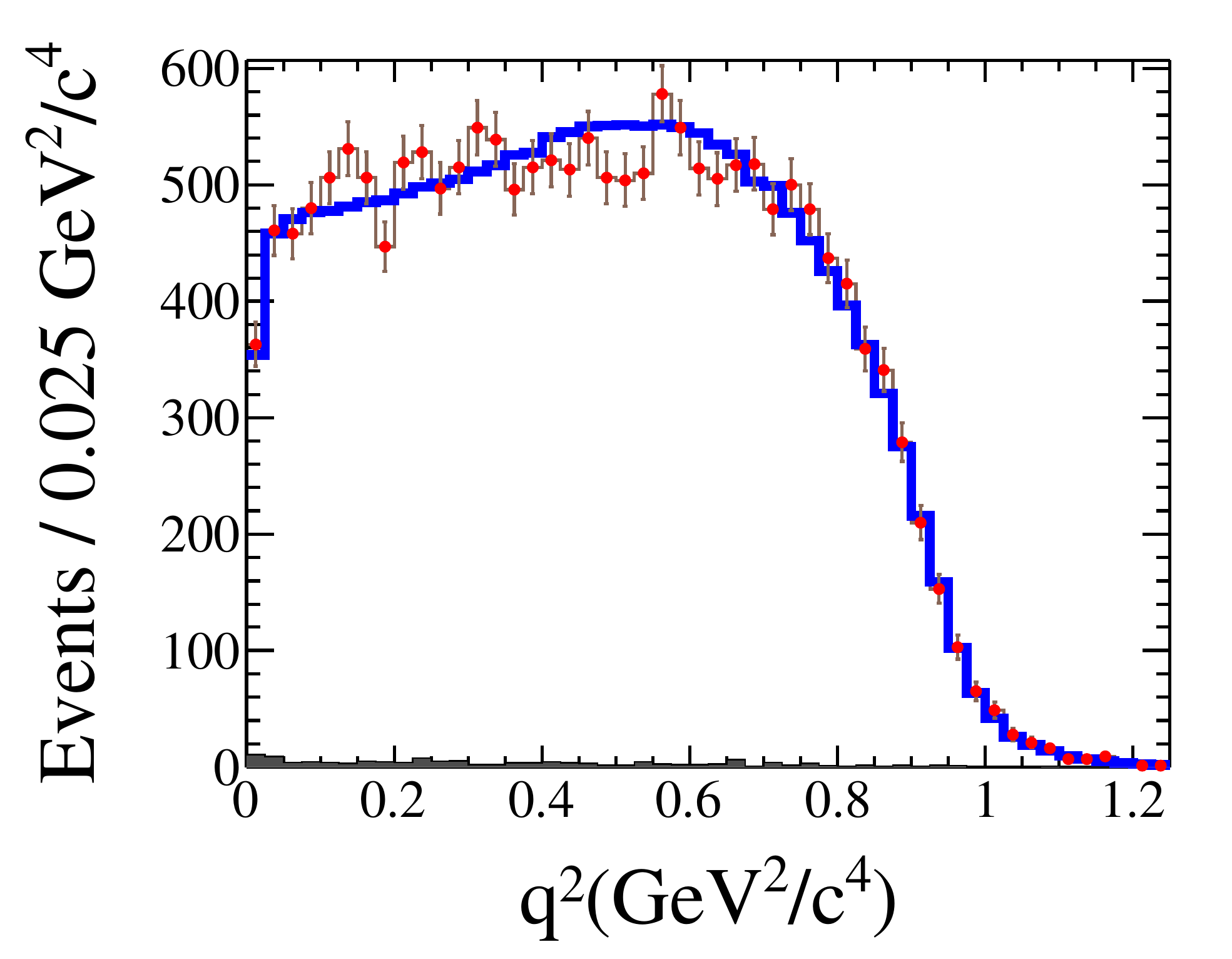} \\
  \includegraphics[width=0.28\linewidth]{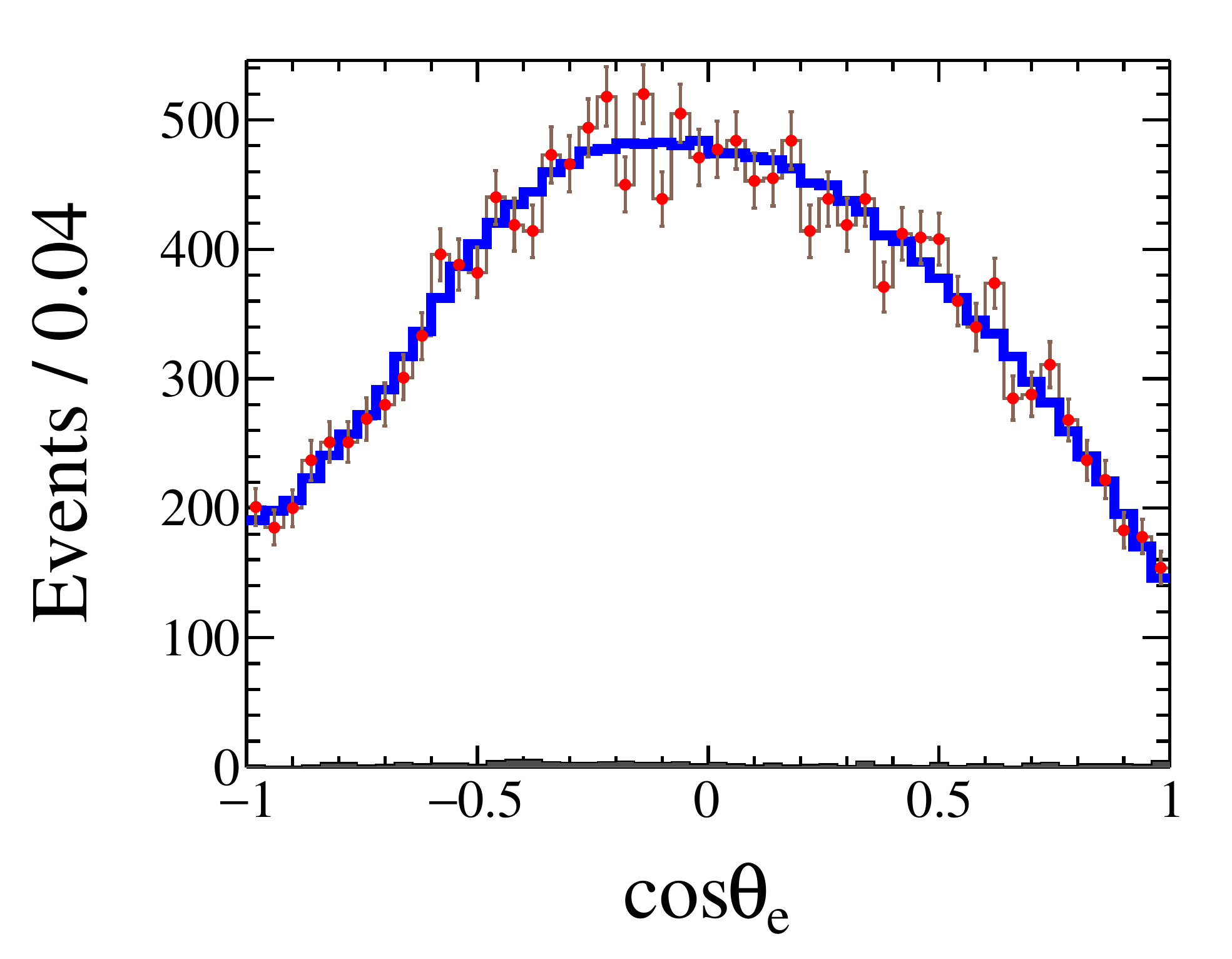} 
  \includegraphics[width=0.28\linewidth]{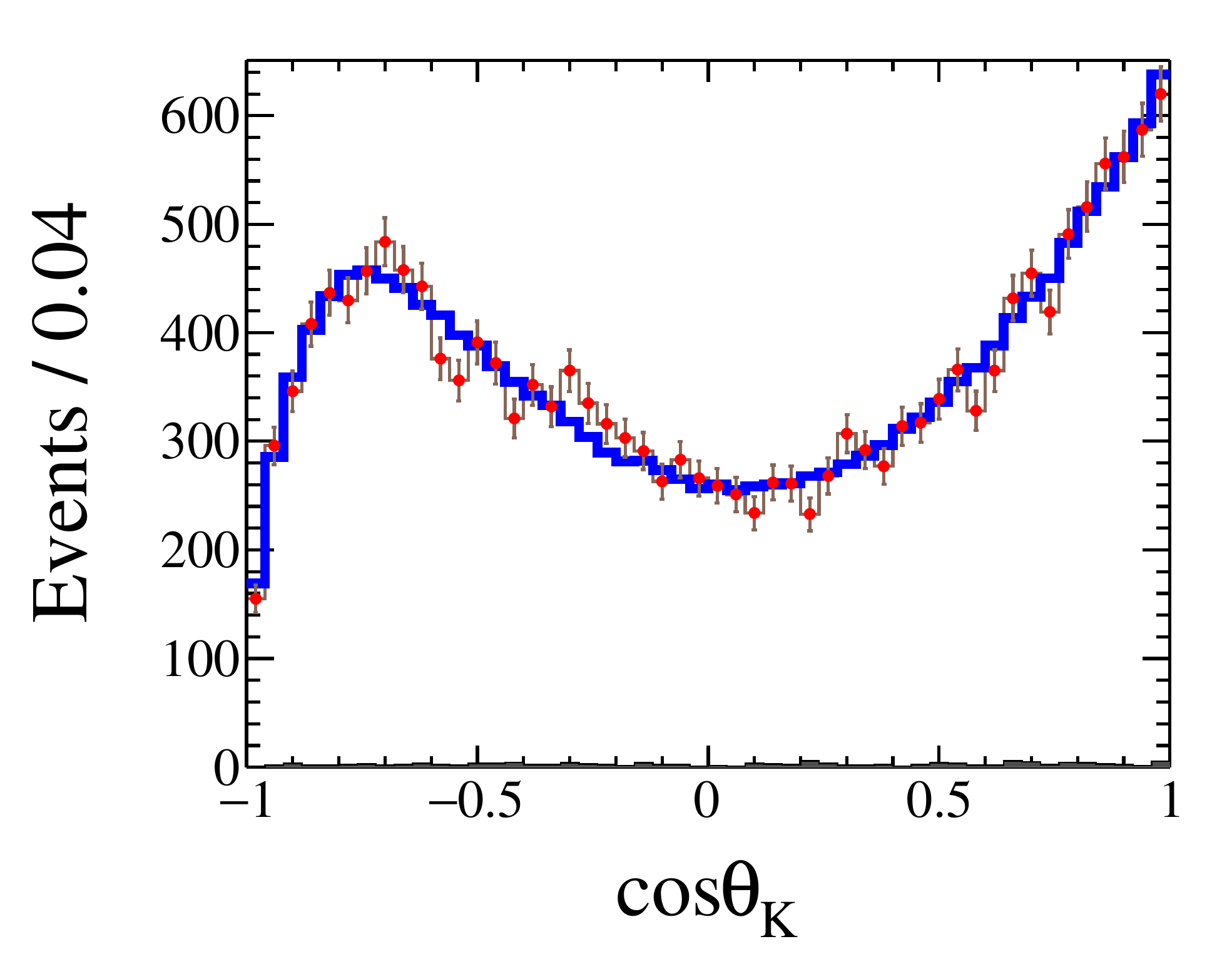}
  \includegraphics[width=0.28\linewidth]{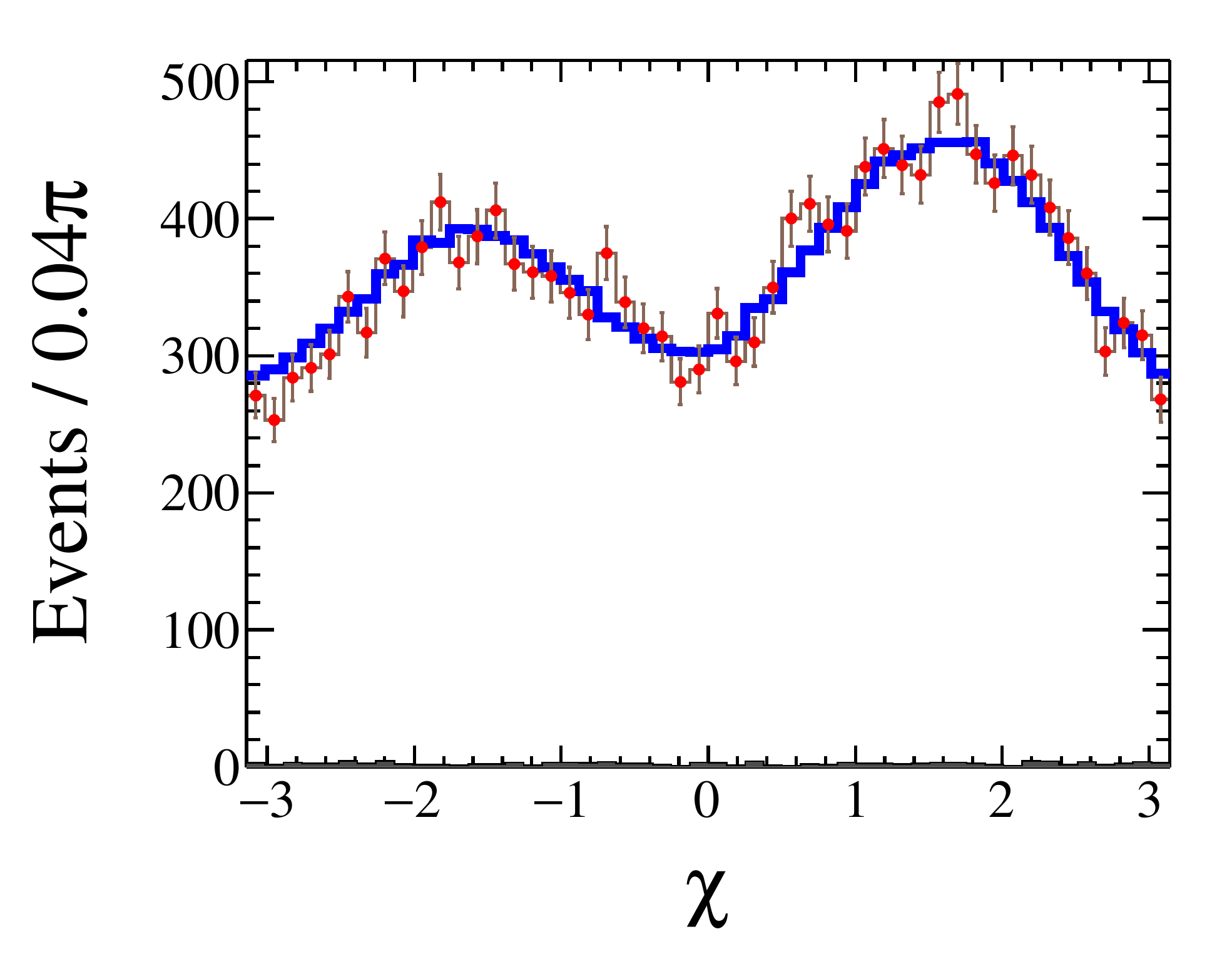}
  \caption{Projections onto each of the kinematic variables, 
  comparing data (dots with error bars) and signal MC 
  weighted by PWA solution (line),
  assuming that the signal is composed of the \emph{S}-wave and $\bar K^{*}(892)^{0}$. 
  The shadowed histogram is the estimated background.}
  \label{fig:projection_SP}
  \end{center}
\end{figure}

The PWA shows that besides the dominant $\bar K^{*}(892)^{0}$,
an $S$-wave contribution making of the non-resonant background term 
and the $\bar K^{*}_{0}(1430)^{0}$ accounts for
 $(6.05\pm0.22\pm0.18)$\%. Other components can be neglected.
 The $\bar K^{*}(892)^{0}$ parameters
are determined: 
$m_{\bar K^{*}(892)^{0}}=(894.60\pm0.25\pm0.08)~ \rm MeV/c^2$,
$\Gamma_{\bar K^{*}(892)^{0}}=(46.42\pm0.56\pm0.15)~ \rm MeV/c^2$,
and the Blatt-Weisskopf parameter 
$r_{BW}=3.07\pm0.26\pm0.11 ~(\rm GeV/c)^{-1}$.  The parameters defining the
hadronic form factors are also measured:
$r_{V} = \frac{V(0)}{A_{1}(0)} = 1.411\pm0.058\pm0.007$,
$r_{2} = \frac{A_{2}(0)}{A_{1}(0)} = 0.788\pm0.042\pm0.008$,
$m_{V} = (1.81^{+0.25}_{-0.17}\pm0.02)~ \rm MeV/c^{2}$,
$m_{A} = (2.61^{+0.22}_{-0.17}\pm0.03)~ \rm MeV/c^{2}$,
$A_{1}(0) = 0.585\pm0.011\pm0.017$.
$m_{V}$ is firstly measured for this decay. Corresponding
projections over the five kinematic variables 
are illustrated in Figure~\ref{fig:projection_SP}.

\begin{figure}[h]
  \centering
  \includegraphics[width=0.35\linewidth]{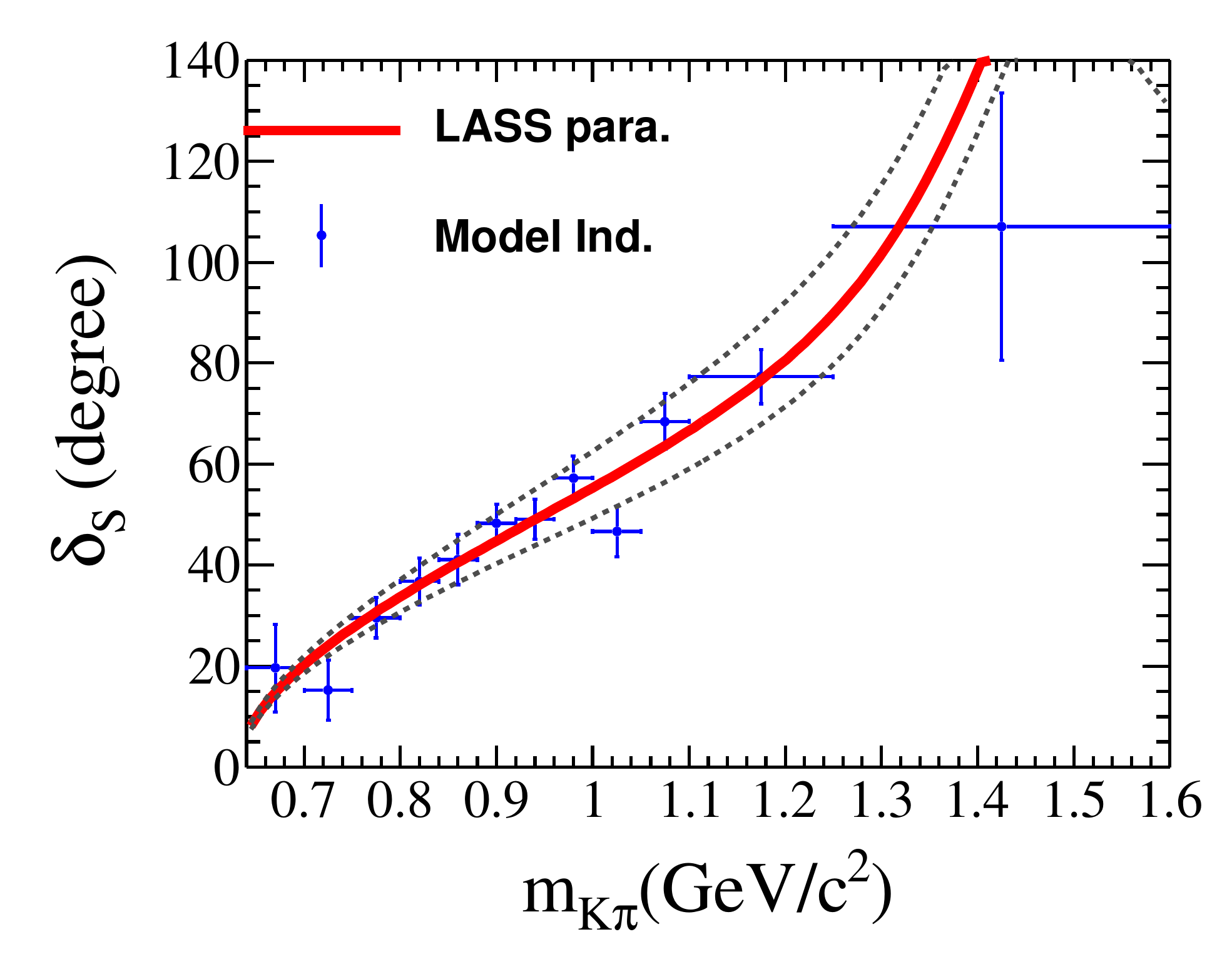}
  \caption{The \emph S-wave phase variation versus $m_{K\pi}$
  assuming that the signal is composed of the
  \emph S-wave and  $\bar K^{*}(892)^{0}$. 
  Model-independent measurement (points with error bars)
  is compared with the result based 
  on the LASS parameterization (solid line, 
  1$\sigma$ deviation is marked by dashed line).}
  \label{fig:pwa_Sphase}
\end{figure}

In the above PWA process $\delta_{S}$ depends on 
$m_{K\pi}$ according to the LASS parameterization.
Then we measure $\delta_{S}$ in a model-independent way. 
We divide the $m_{K\pi}$ spectrum into twelve bins
and perform the PWA fit with  $\delta_{S}$
in each bin as twelve additional fit parameters (within each bin 
the phase is assumed to be constant).
Figure~\ref{fig:pwa_Sphase} illustrates the comparison of 
the model-independent measurement with that 
based on the LASS parameterization.

In the PWA, the helicity  basis form factors are 
assumed to depend on $q^2$ according to the spectroscopic pole dominance (SPD) model.
In order to achieve a better understanding of the 
semileptonic decay dynamics, we also measure the 
 form factors in a model-independent way using
the projective weighting technique, which is introduced 
in Ref.~\cite{Link:2005dp}.

\begin{figure}[h]
  \includegraphics[width=0.28\linewidth]{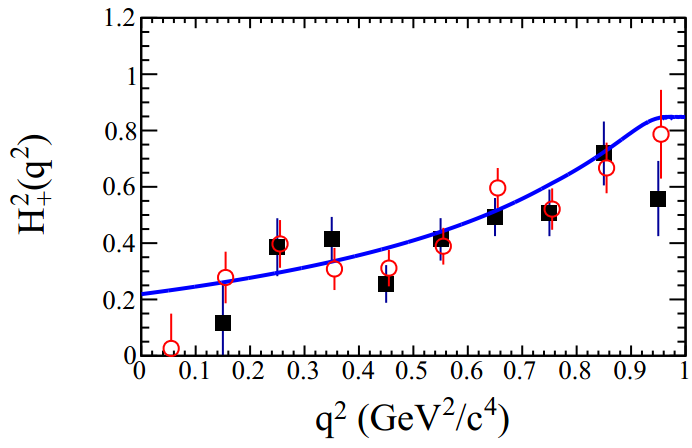}
  \includegraphics[width=0.28\linewidth]{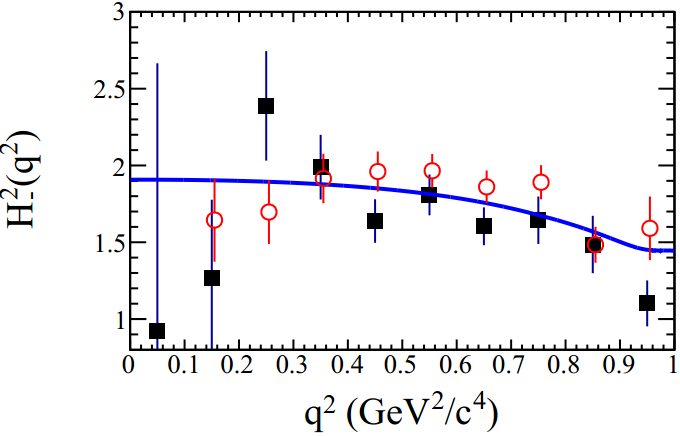}
  \includegraphics[width=0.28\linewidth]{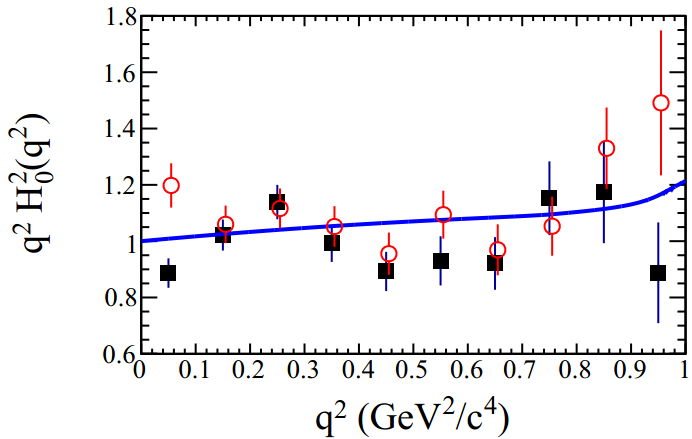} \\
  \includegraphics[width=0.28\linewidth]{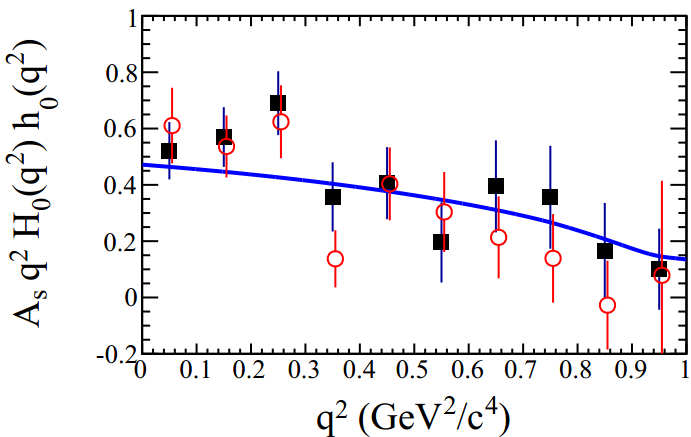}
  \includegraphics[width=0.28\linewidth]{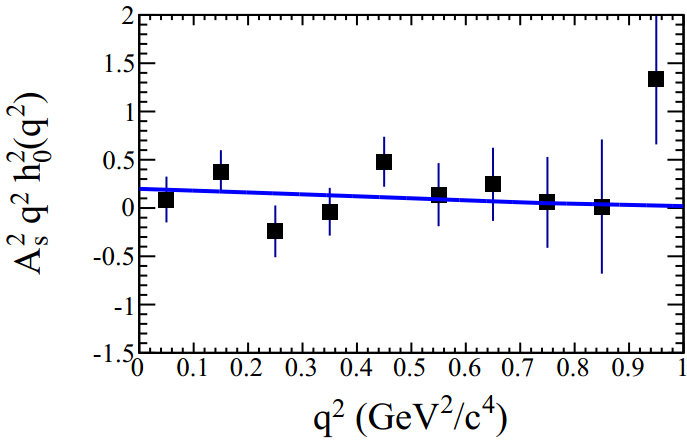}
  \caption{Form factors measured in this work (squares) 
compared with the CLEO-c 
results (circles) and with the PWA solution (curves).
  Error bars represent both statistical and systematic uncertainties.}
  \label{fig:pwt}
\end{figure}

Only candidates in the $\bar{K}^{*}(892)^{0}$ region
([0.8,1] GeV/c$^2$) are used, so that we can 
neglect other components than the $\overline{K}^{*}(892)^{0}$ and 
the non-resonant $S$-wave. The
decay intensity of $D^{+}\to K^{-}\pi^{+}e^{+}\nu_{e}$
can be parameterized by form factors describing
the decay into the vector meson $\bar{K}^{*}(892)^{0}$:
 $H_{+}(q^2,m)$,  $H_{-}(q^2,m)$,  $H_{0}(q^2,m)$, 
 and by an additional form factor  $h_{0}(q^2,m)$ describing 
 the non-resonant $S$-wave contribution. 
 The form factors are measured by weighting the $q^2$
  distributions based on the angular. 
  The results are shown in Figure~\ref{fig:pwt}.
  They are consistent with the SPD model with the parameters
obtained from the PWA, and  with the results
reported by CLEO-c~\cite{Briere:2010zc}.

\section{Measurement of the form factors in the decay $D^{+}\to \omega e^{+}\nu_{e}$
and search for the decay $D^{+}\to \phi e^{+}\nu_{e}$}
\label{sec:pipienu}

The decay $D^{+}\to \omega e^{+}\nu_{e}$ has similar dynamics as
$D^{+}\to \bar{K}^{*}(892)^{0} e^{+}\nu_{e}$. Neglecting the mass 
of the electron, the transition matrix element of $D$ to vector meson 
can be decomposed into contributions from one vector 
$V(q^2)$ and  two axial-vector  ($A_{1}$, $A_{2}$)($q^{2}$) form factors,
 where $q^2$ is the 
invariant mass square of the $e^{+}\nu_{e}$ system.
A precise measurement of the branching ratio and the form factors 
provide opportunities to test the standard model and 
the theoretical calculations.

The decay $D^{+}\to \phi e^{+}\nu_{e}$ has not been observed at present.
The $\phi$ ($s\bar{s}$) has different quark composition from 
the $D$ meson ($c\bar{d}$), so the process can only
proceed either through $\omega-\phi$ mixing or 
non-perturbative ``weak annihilation" (WA).
A measurement of the branching fraction can 
discriminate which process is dominant.

\begin{figure}[h]
  \begin{center}
  \includegraphics[width=0.4\linewidth]{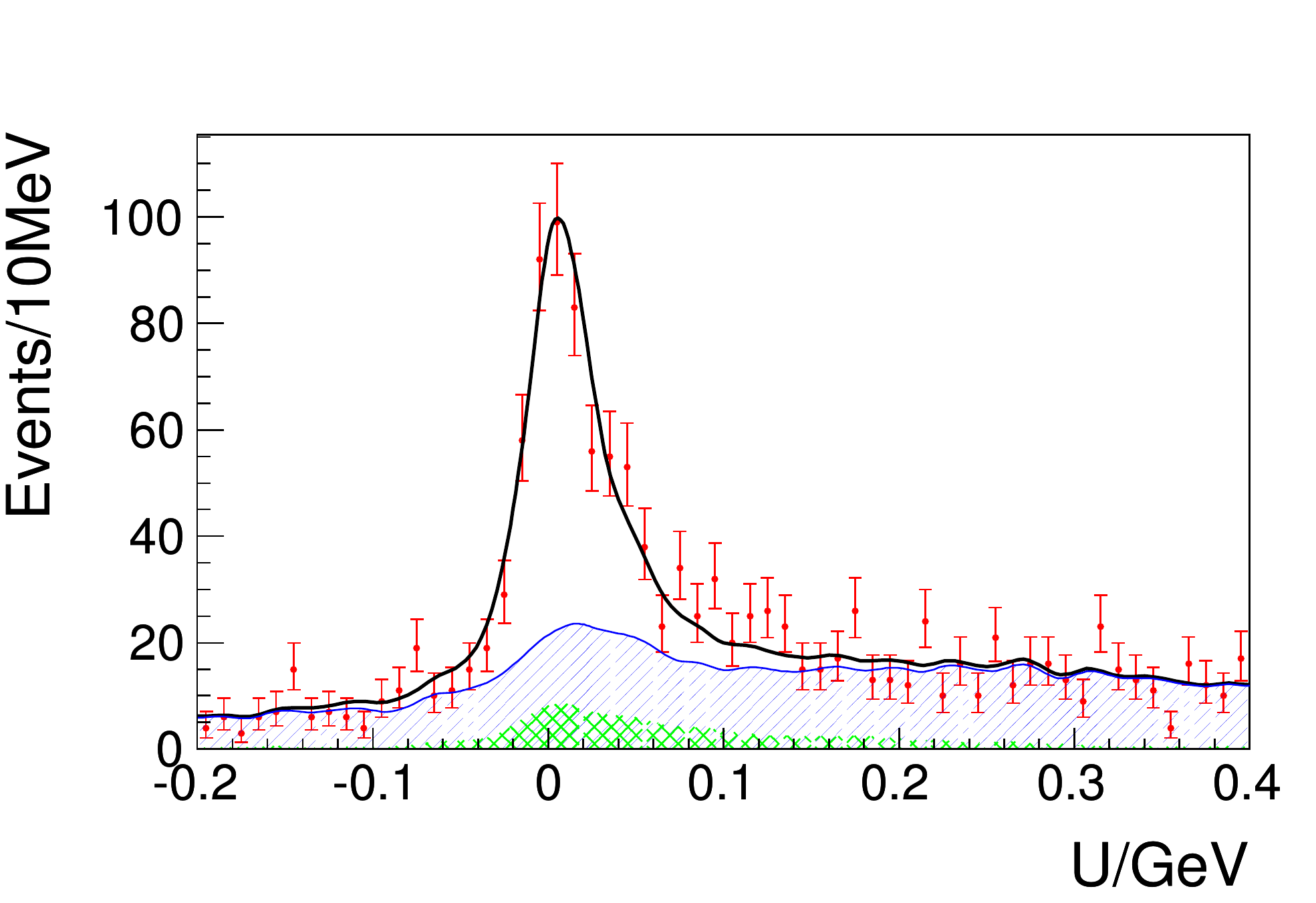}
  \includegraphics[width=0.4\linewidth]{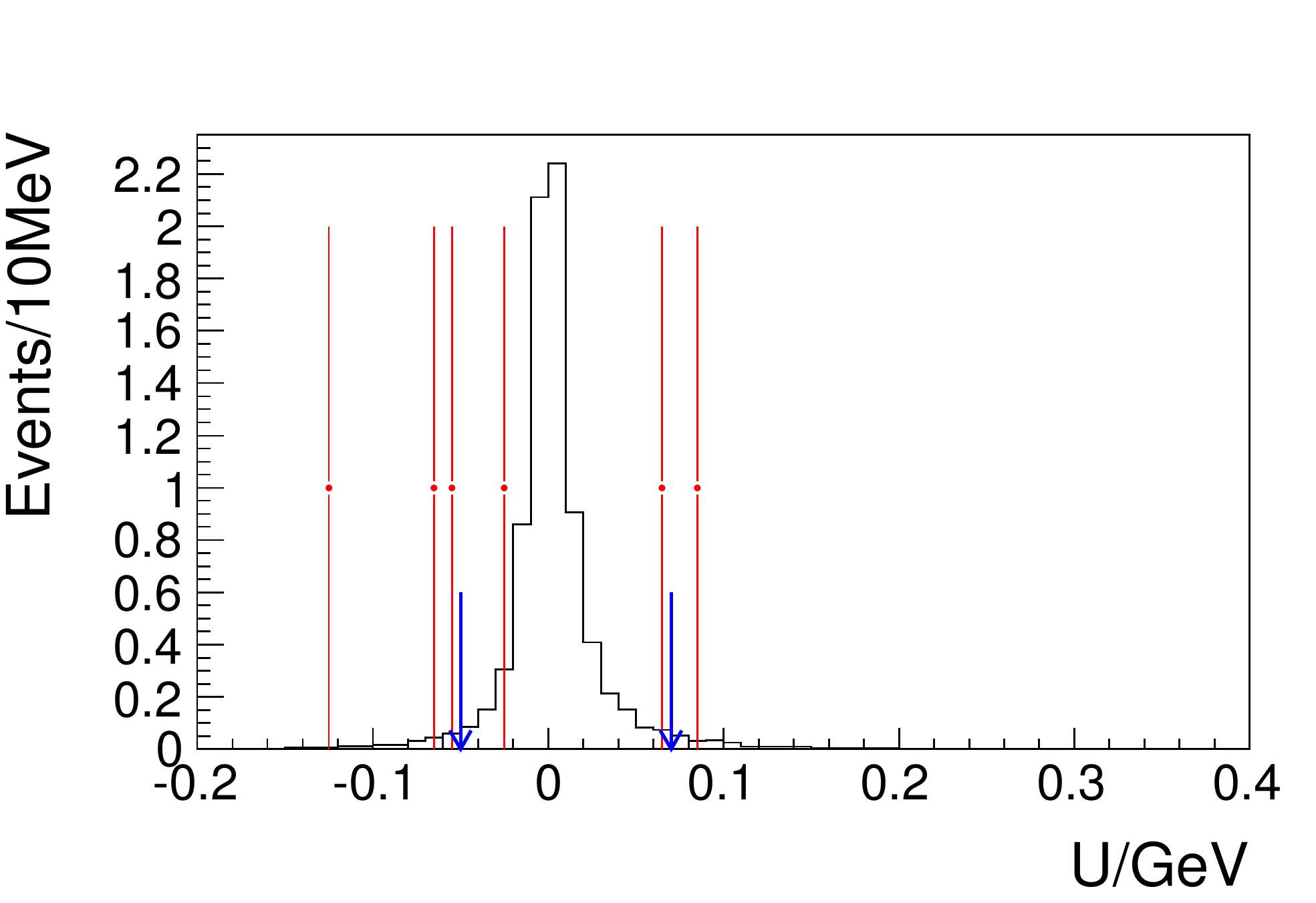}
  \caption{Left: fit (solid line) to the $U$ distribution
  in data (points with error bars) for $D^{+}\to \omega e^{+}\nu_{e}$.
  The total background is shown by the filled curve, 
  with the peaking component shown by the cross-hatched curve.
  Right: the $U$ distribution for $D^{+}\to \phi e^{+}\nu_{e}$
  in data (points with error bars) and signal MC with 
  arbitrary normalization (solid histograms). The arrows 
  show the signal region.}
  \label{fig:pipienu_signal}
  \end{center}
\end{figure}

The signal yields are determined by the variable 
$U$, the difference between the missing energy and 
momentum~\cite{CLEO:2011ab}.
The yield of $D^{+}\to \omega e^{+}\nu_{e}$ is obtained 
from a fit to the $U$ distribution as shown in the 
left plot of Figure~\ref{fig:pipienu_signal}.
And that of $D^{+}\to \phi e^{+}\nu_{e}$ is obtained 
by counting the number in the signal region
[-0.05, 0.07] GeV as shown in the right plot,
indicating no significant excess of signal events. 
The results are concluded in Table~\ref{tab:pipienu},
which are improved compared with previous reports.

\begin{table}[h]
\small
\begin{center}
\caption{Measured branching fractions and a comparison 
with the previous measurements.
For $D^{+}\to \omega e^{+}\nu_{e}$, the first uncertainty 
is statistical and the second systematic.}
\begin{tabular}{l|cc}  
Mode &  This work &  Previous~\cite{CLEO:2011ab,Yelton:2010js}  \\ \hline
$\omega e^{+}\nu_{e}$  &   $(1.63\pm0.11\pm0.08)\times 10^{-3}$ 
    &   $(1.82\pm0.18\pm0.07)\times 10^{-3}$    \\
$\phi e^{+}\nu_{e}$    &  $1.3\times10^{-5}$ (90\%C.L.)    
  &     $9.0\times10^{-5}$ (90\%C.L.) \\ \hline
\end{tabular}
\label{tab:pipienu}
\end{center}
\end{table}

\begin{figure}[h]
  \begin{center}
  \includegraphics[width=0.82\linewidth]{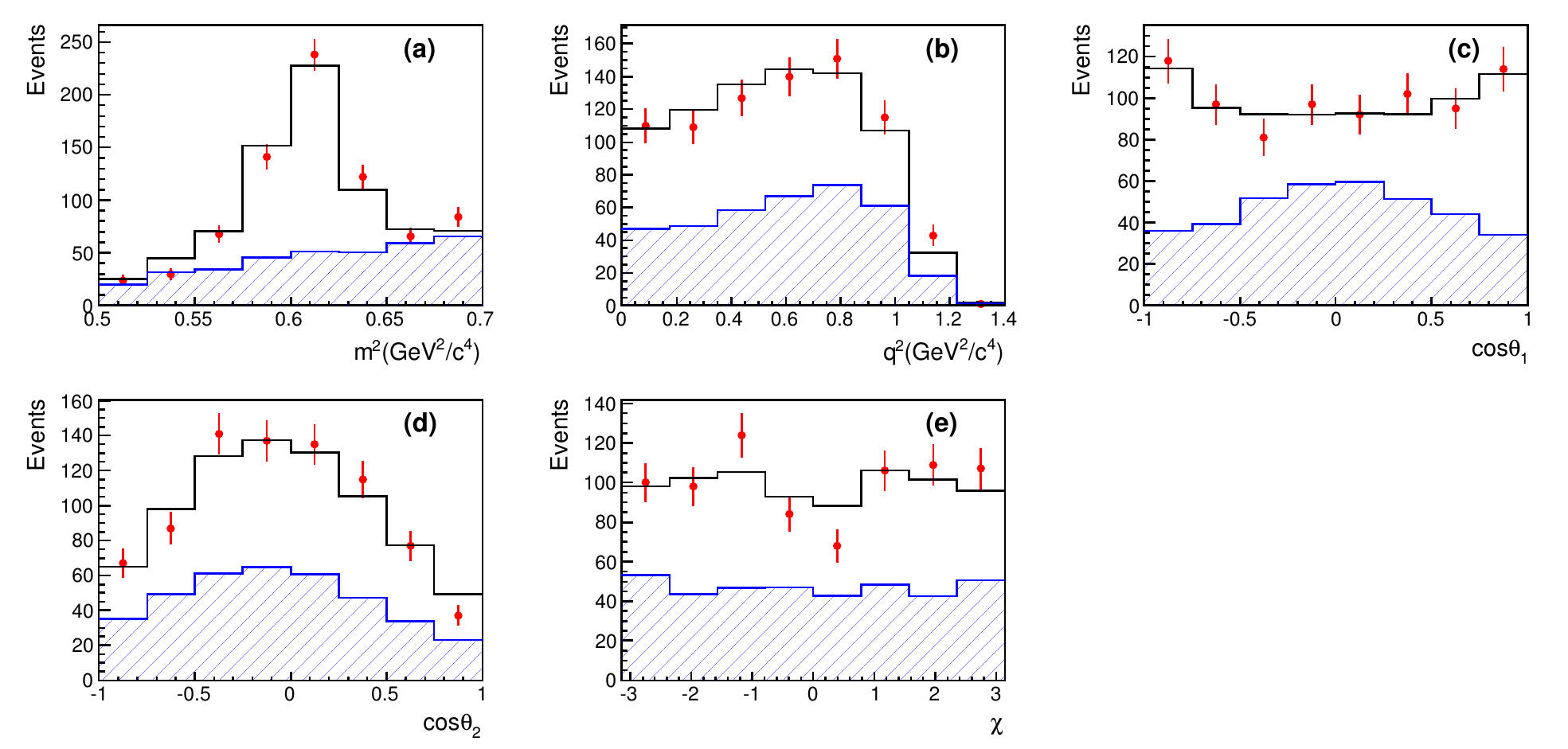}
  \caption{Projections of the data set (points with error bars), the fit results
  (solid histograms) and the sum of the 
  background distributions (filled histogram curves) onto
  (a) $m^2$, (b) $q^2$, (c) cos$\theta_{1}$, (d) cos$\theta_{2}$ 
  and (e) $\chi$.}
  \label{fig:pipienu_ff}
  \end{center}
\end{figure}

In order to measure the form factors in the decay
$D^{+}\to \omega e^{+}\nu_{e}$, a five-dimensional 
maximum likelihood fit is performed in the space of 
$m^2$, $q^2$,  cos$\theta_{1}$, cos$\theta_{2}$ and $\chi$,
whose definitions are similar with those
 described for the decay $D^{+}\to K \pi e^{+}\nu_{e}$
  in Sec.\ref{sec:kpienu}.
  The form factor ratios are determined from the fit:
  $r_{V}=1.24\pm0.09\pm0.06$, $r_{2}=1.06\pm0.15\pm0.05$,
  which are measured for the first time in this decay.
  The fitted projections over the five variables are 
  illustrated in Figure~\ref{fig:pipienu_ff}.

\section{Study of decay dynamics and \emph{CP} asymmetry in $D^{+}\to K^{0}_{L} e^{+}\nu_{e}$ decay}
\label{sec:kenu}

In charged $D$ meson decays, $CP$ asymmetry occurs when the absolute 
value of the decay amplitude for $D$ decaying to a final state is
different from the one for the corresponding $CP$-conjugated  amplitude.
The most optimistic model-dependent estimates put the SM predictions 
for the asymmetry as $\mathcal{O}(10^{-3})$ or below~\cite{Buccella:1994nf},
so an observation of any $CP$-violating signal would  be a sign of new
physics. 
The hadronic matrix element of $D$ to pseudoscalar meson process 
can be decomposed into contributions from longitudinal and  
transverse form factors. Neglecting the lepton mass,
only the transverse form factor contributes. Various 
theoretical techniques provide slightly different $q^2$
dependencies of the form factors, High precision measurements
of the partial decay width over different ranges of $q^2$
will distinguish which method correctly describes the 
non-perturbative dynamics of QCD. The $D^{+}\to K^{0}_{L} e^{+}\nu_{e}$
decay is investigated in this work with its 
branching fraction and $CP$ violation firstly measured.
$q^2$ dependence of the form factors are measured  based on
different theoretical models for the first time as well.

In the process of searching for $D^{+}\to K^{0}_{L} e^{+}\nu_{e}$ signals,
the basic idea for $K^{0}_{L}$ search is explicitly introduced. 
Most $K^{0}_{L}$ will penetrate the MDC,  interact with 
the scintillators in the EMC and provide the position information. 
With all charged particles in the event reconstructed and applying 
energy-momentum conservation to all the particles,
the direction of the missing $K^{0}_{L}$ can be determined.
We then examine whether there's a good neutral shower deposited
in the EMC along the predicted $K^{0}_{L}$  direction.

\begin{figure}[h]
  \begin{center}
  \includegraphics[width=0.7\linewidth]{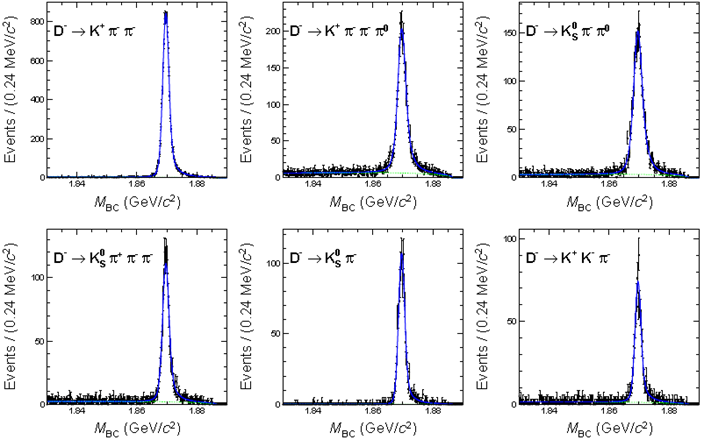}
  \caption{Fits to the $M_{BC}$ distributions of the tagged $D^{-}$ 
  candidates for data, requiring $D^{+}\to K^{0}_{L} e^{+}\nu_{e}$. 
  The dots with error bars are for data, and the blue solid curves
  are the results of the fits. The green dashed curves are the fitted 
  backgrounds.}
  \label{fig:kenu_signal}
  \end{center}
\end{figure}

The signal yield is obtained by fitting the distributions of the
beam constrained mass ($M_{BC}$) of the corresponding tagged $D$ candidates,
as illustrated in Figure~\ref{fig:kenu_signal}.
Branching fractions of the six tag modes are
measured separately for $D^{+}$ and $D^{-}$.
Then the weighted branching fractions for $D^{+}$ and $D^{-}$ are obtained:
$\mathcal{B}(D^{+}\to K^{0}_{L} e^{+}\nu_{e}) = (4.454\pm0.038\pm0.102)\%$,
$\mathcal{B}(D^{-}\to K^{0}_{L} e^{-}\bar{\nu}_{e}) = (4.507\pm0.038\pm0.104)\%$.
The combined averaged value is
\begin{equation}
\nonumber
\mathcal{B}(D^{+}\to K^{0}_{L} e^{+}\nu_{e}) = (4.481\pm0.027\pm0.103)\%,
\nonumber
\end{equation}
which agrees well with the measurement of 
$\mathcal{B}(D^{+}\to K^{0}_{S} e^{+}\nu_{e})$ by CLEO-c~\cite{Besson:2009uv}.
The $CP$ asymmetry is determined to be
\begin{equation}
A_{CP} \equiv \frac{\mathcal{B}(D^{+}\to K^{0}_{L} e^{+}\nu_{e})-\mathcal{B}(D^{-}\to K^{0}_{L} e^{-}\bar{\nu}_{e})}
{\mathcal{B}(D^{+}\to K^{0}_{L} e^{+}\nu_{e})+\mathcal{B}(D^{-}\to K^{0}_{L} e^{-}\bar{\nu}_{e})}
=(-0.59\pm0.60\pm1.48)\%,
\end{equation}
which is consistent with the theoretical prediction in 
Ref.~\cite{Xing:1995jg}.

\begin{figure}[h]
  \begin{center}
  \includegraphics[width=0.72\linewidth]{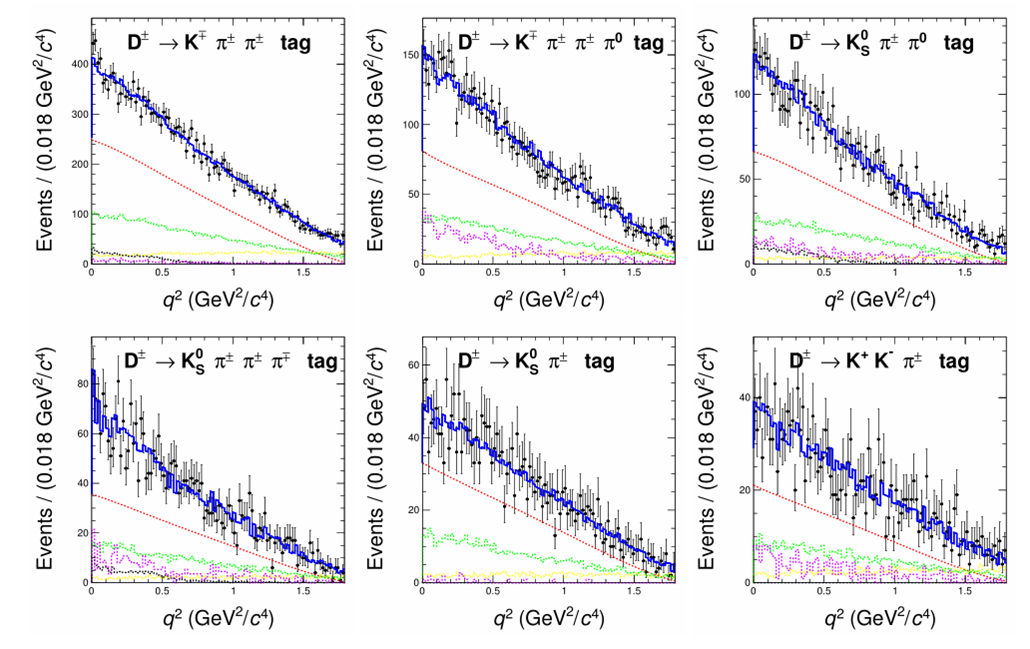}
  \caption{(Color online) Simultaneous fits (blue solid curves) 
  to the numbers of   $D^{+}\to K^{0}_{L} e^{+}\nu_{e}$ data 
  (points with error bars) as a function of $q^2$
  with the two-parameter series expansion parameterization. 
  The red dashed curves show the signal, while
  the violet, yellow, green and black curves 
  refer to different kinds of backgrounds. } 
  \label{fig:kenu_ff}
  \end{center}
\end{figure}

We perform simultaneous fits to the distributions of the observed
 $D^{+}\to K^{0}_{L} e^{+}\nu_{e}$ candidates as a function of $q^2$
to determined the product of the hadronic form 
factor and the CKM matrix element $f_{+}^{K}(0)|V_{cs}|$.
The form-factor shape is described based on the simple pole model ($m_{pole}$),
the modified pole model ($\alpha$), two-parameter series expansion 
($r_{1}$), and three-parameter series expansion ($r_{1}$, $r_{2}$). 
As an example, Figure~\ref{fig:kenu_ff} shows the simultaneous
 fits using the two-parameter series expansion model,
corresponding to the results: $f_{+}^{K}(0)|V_{cs}|=0.728\pm0.006\pm0.011$,
 $r_{1}=-1.91\pm0.33\pm0.24$.
 
\section{Summary}

Based on 2.92 fb$^{-1}$ of data collected by the BESIII experiment three 
semileptonic decays are analysed. In the study of 
$D^{+}\to K^{-}\pi^{+}e^{+}\nu_{e}$, its branching fractions 
are measured: $\mathcal{B}(D^{+}\to K^{-}\pi^{+}e^{+}\nu_{e})=(3.71\pm0.03\pm0.08)\%$,
$\mathcal{B}(D^{+}\to K^{-}\pi^{+}e^{+}\nu_{e})_{[0.8,1]}=(3.33 \pm 0.03 \pm 0.07)\%$.
A PWA is performed and an \emph{S}-wave contribution is found to account for 
($6.05\pm0.22\pm0.18$)\%. The \emph{S}-wave phase and the form factors are measured 
both by the PWA and in a model-independent way, showing good consistency.
For the decay $D^{+}\to \omega e^{+}\nu_{e}$,
 the branching fraction is measured with a higher precision: 
 $\mathcal{B}(D^{+}\to \omega e^{+}\nu_{e})=(1.63\pm0.11\pm0.08)\%$.
 Its form factors are determined for the first time:
 $r_{V}=1.24\pm0.09\pm0.06$,   $r_{2}=1.06\pm0.15\pm0.05$.
 The rare decay $D^{+}\to \phi e^{+}\nu_{e}$ is searched for and
  an upper limit  at 90\% confidence level is set:
  $\mathcal{B}(D^{+}\to \phi e^{+}\nu_{e}) < 1.3\times 10^{-5}$.
For the decay $D^{+}\to K_{L} e^{+}\nu_{e}$,
we present the first measurement of the absolute branching fraction
$\mathcal{B}(D^{+}\to K_{L}^{0} e^{+}\nu_{e})=(4.481\pm0.027\pm0.103)\%$.
The $CP$ asymmetry is also determined 
$A_{CP} = (-0.59\pm0.60\pm1.48)\%$, indicating no significant $CP$ violation.
The product $f_{+}^{K}(0)|V_{cs}|$ is measured based on three theoretical form-factor models.
 Taking the two-parameter series expansion parameterization,
  $f_{+}^{K}(0)|V_{cs}|=0.748\pm0.007\pm0.012$.

\Acknowledgements

This work is supported in part by National Natural Science Foundation of China (NSFC) under Contracts Nos. 11075174, 11475185.


\end{document}